\documentclass[12pt]{article}

\usepackage{amsfonts}
\usepackage{amssymb}
\usepackage{amscd}

\def\D{\hbox{D\kern-.73em\raise.25ex\hbox{-}\raise-.25ex\hbox{ }}}
 \def\d{\hbox{d\kern-.33em\raise.75ex\hbox{-}\raise-.75ex\hbox{}}}

\def\GGG{\frak G }
\def\gr3{\GGG\,(\SSS_3)}

\def\gr2{\GGG\,(\SSS_2)}

\def\SSS{\frak S}

\def\ed{\end{document}}
\def\beq{\begin{equation}}
\def\eeq{\end{equation}}
\def\bea{\begin{eqnarray}}
\def\eea{\end{eqnarray}}
\def\ba{\begin{array}}
\def\ea{\end{array}}
\def\bi{\begin{itemize}}
\def\ei{\end{itemize}}

\def\noi{\noindent}

\newcommand{\bp}{\noindent\begin{minipage}[c]}
\newcommand{\ep}{\end{minipage}}

\begin{document}
 \baselineskip=11pt

\title{\bf \Large  $p$-Adic and Adelic \\
  Rational Dynamical Systems\hspace{.25mm} }
\author{{Branko Dragovich}\hspace{.25mm}\thanks{\,e-mail
address: dragovich@phy.bg.ac.yu}
\\ \normalsize{Institute of Physics, P.O. Box 57, 11001 Belgrade, Serbia}
\vspace{2mm}
\\
{Du\v san Mihajlovi\'c}\hspace{.25mm}
\\ \normalsize{Faculty of Physics, P.O. Box 368, 11001 Belgrade, Serbia}}

\date{}

\maketitle


\begin{abstract}
In the framework of adelic approach we  consider real and $p$-adic
properties of dynamical system given by linear fractional map $f (x)
= (a \, x + b)/(c x + d)$, where $a ,\, b,\, c,$ and $d$ are
rational numbers.  In particular, we investigate behavior of this
adelic dynamical system when fixed points are rational. It is shown
that any of  rational fixed points is $p$-adic indifferent for all
but a finite set of primes. Only for finite number of $p$-adic cases
a rational fixed point may be attractive or repelling. The present
analysis is a continuation of the paper math-ph/0612058. Some
possible generalizations are discussed.

\end{abstract}

\section{\large Introduction }

Many dynamical systems  change their states in discrete time
intervals by a mapping
 \beq         f : X \rightarrow X ,  \label{1.1} \eeq
where $X$ is the space of states and  $f$ describes how states $x
\in X$ evolve in time. If the state at the time $t=0$ is $x_{0} \in
X$ and $f^{n}= f \circ \cdots \circ f$ then after $n$ iterations the
state becomes
\beq x_{n} = f^{n}(x_{0}) . \label{1.2}\eeq

$X$ has usually some natural  structures, e.g.  hierarchies and
distances between states. In physics of very complex systems $X$
often displays a hierarchical structure, which implies that the
classification of the states and their relationships should use
ultrametric distances, and in particular $p$-adic ones. Recently
much attention has been paid to some $p$-adic dynamical systems,
since they have a lot of potential applications (for a review, see
\cite{khrennikov4}).

Ground states of the mean field models for spin glasses  have
ultrametric structure  \cite{parisi1}. Methods of p-adic analysis
are applied to the investigation of replica symmetry breaking
\cite{parisi2} and p-adic reformulation of  the ultrametric
structure of spin glasses \cite{khrennikov2}.

During the last two decades there have been many constructions of
$p$-adic physical models. In particular, $p$-adic numbers have been
successfully used in string theory, quantum mechanics and quantum
cosmology (for a review, see \cite{freund}, \cite{vladimirov},
\cite{dragovich1} and \cite{dragovich2}).

Presently it is not known any physical principle or phenomenon that
would point out a particular prime number. Moreover, mathematical
objects, e.g. such as the Riemann zeta function, are very
significant when all primes are employed on the equal footing (see
\cite{dragovich} for a recent example). Simultaneous  use of the
real and $p$-adic numbers, which make all possible completions of
the field $\mathbb{Q}$ of rational numbers, is also of great
importance in mathematics. Their use in the form of adeles is
particularly effective in the arithmetic theory of algebraic groups.
Adelic models of physical  systems contain real and $p$-adic
submodels as parts of a whole (see, e.g. \cite{dragovich3}). They
give more information on a dynamical system than real and $p$-adic
treatments separately. Since 1987 adelic models have been
constructed and investigated in string theory, quantum mechanics,
quantum cosmology (for a review, see \cite{freund},
\cite{vladimirov}, \cite{dragovich1} and \cite{dragovich2}) and in
some other fields of modern mathematical physics (see, e.g.
\cite{karwowski}).

In the recent article \cite{mihajlovic} we started  $p$-adic and
adelic investigation of dynamical systems, which evolution is
governed  by linear fractional transformations
 \beq
 f (x) = \frac{a x + b}{c x + d}\,,   \label{1.3}
\eeq \noi where $a, b, c, d \in \mathbb{Q}$ with conditions $x \neq
-\frac{d}{c}$, $c \neq 0$ and $a d - b c = 1$.

Some $p$-adic properties of this kind of dynamical systems were
explored in \cite{mukhamedov1}, where parameters $a\,, b\,, c\,, d
\in \mathbb{C}_p$. It is worth noting that taking  physical
parameters to be rational numbers gives a possibility to treat real
and $p$-adic properties simultaneously and on the equal footing.

Linear fractional transformations (M\"obius transformations)
(\ref{1.3}) and related  $S L (2, \mathbb{C})$, $S L (2,
\mathbb{C}_p)$ groups, and their subgroups, have very rich
mathematical structures. They also have important applications in
many parts of mathematical and theoretical physics (see, e.g.
\cite{freund}, \cite{todorov} \cite{dolan} and references therein).

Sec. 2 contains a very brief introductory  review of $p$-adic
numbers and adeles. In Sec. 3  some new results of the above linear
fractional dynamics (\ref{1.3}) are presented. Some general remarks,
including possible generalizations, are stated in Sec. 4.

\section{\large $p$-Adic Numbers and Adeles}

Rational numbers are significant in physics as well as in
mathematics. Physical significance comes from the fact that a result
of any measurement is a rational number. One can obtain the field
${\mathbb R}$ of real numbers from ${\mathbb Q}$ by employing  the
absolute value, which is an example of the norm (valuation) on
${\mathbb Q}$.  In addition to the absolute value, for which  we use
usual arithmetic notation $|\cdot|_\infty$, one can introduce on
${\mathbb Q}$ a norm with respect to each prime number $p$. Note
that any rational number can be uniquely written as $x = p^\nu \,
\frac{m}{n}$, where $p,\, m,\, n$ are mutually prime and $\nu \in
{\mathbb Z}$. Then by definition $p$-adic norm (or, in other words,
$p$-adic absolute value) is $| x|_p = p^{-\nu}$ if $ x \neq 0$ and
$|0|_p =0$. One can verify that $|\cdot |_p$ satisfies the strong
triangle inequality, i.e. $|x + y|_p \leq \, \mbox{max} \, (|x|_p
\,,\, |y|_p) $. Thus $p$-adic norms belong to the class of
non-Archimedean (ultrametric) norms.  According to the Ostrowski
theorem any nontrivial norm on ${\mathbb Q}$ is equivalent either to
the $|\cdot|_\infty$ or to one of the $|\cdot|_p$. One can easily
show that $|m|_p \leq 1$ for any $m \in {\mathbb Z}$ and any prime
$p$. The $p$-adic norm is a measure of divisibility of the integer
$m$ by prime $p$: the more divisible, the $p$-adic smaller. Using
Cauchy sequences of rational numbers one can make completions of
${\mathbb Q}$ to obtain ${\mathbb R} \equiv {\mathbb Q}_\infty$ and
the fields ${\mathbb Q}_p$ of $p$-adic numbers using norms
$|\cdot|_\infty$ and $|\cdot|_p \,$, respectively.  $p$-Adic
completion of ${\mathbb N}$ gives the ring ${\mathbb Z}_p = \{ x \in
{\mathbb Q}_p :\,\, |x|_p \leq 1 \}$ of $p$-adic integers. Denote by
${\mathbb U}_p = \{ x \in {\mathbb Q}_p :\,\, |x|_p = 1 \}$
multiplicative group of $p$-adic units.

Any $p$-adic number $ x \in {\mathbb Q}_p$ can be presented in the
unique way (unlike real numbers) as the sum of $p$-adic convergent
series of the form
\begin{equation}
x = p^\nu \, (x_0 + x_1 p  + \cdots + x_n p^n + \cdots ) , \quad \nu
\in {\mathbb Z} , \quad x_n \in \{0, 1, \cdots, p-1 \} . \label{2.1}
\end{equation}
 If $\nu \geq 0$ in (\ref{2.1}), then $x \in {\mathbb Z}_p$\,. When
$\nu = 0$ and $x_0 \neq 0$ one has $x \in {\mathbb U}_p$\,.

 $p$-Adic metric $d_p (x, y) = |x - y|_p $  satisfies all
necessary properties of metric with strong triangle inequality, i.e.
$d_p (x, y) \leq \mbox{max} \,(\, d_p (x,z), \, d_p (z, y)\,)$ which
is of the non-Archimedean (ultrametric) form. Using this metric,
${\mathbb Q}_p$ becomes an ultrametric space with $p$-adic topology.
A closed  $p$-adic  ball (disk) is  $B_p (r, \, \xi) = \{ x \in
\mathbb{Q}_p \, : \, |x - \xi|_p \leq r \}$, where  $r = p^m  ,\,
m\in \mathbb{Z}\,, $ is  radius with discrete values, and $\xi$ is a
center of the ball. Analogously,  an open ball (disk) is $B_p^- (r ,
\, \xi) = \{ x \in \mathbb{Q}_p \, : \, |x - \xi|_p < r \}$. Sphere
of radius $\rho$ and center $\xi$ is $S_p (\rho, \, \xi) = \{ x \in
\mathbb{Q}_p \, : \, |x - \xi|_p = \rho\}$. Any ball can be regarded
as closed as open. Any point  $ x \in B_p (r , \xi )$ can be treated
as center of the same ball. Note the following connections: $ S_p
(\rho, \, \xi) = B_p (\rho , \, \xi)\setminus B_p^- (\rho , \,
\xi)\,, \,\,\, B_p (r , \, \xi) = \bigcup_{\rho \leq r} S_p (\rho,
\, \xi)$.

It is worth noting that $x \in S_p (\rho, \, \xi)$ has the form $x =
\xi + y = p^k \, (\xi_0 + \xi_1 \, p + \xi_2 \, p^2 + \cdots) + p^l
\, (y_0 + y_1 \, p + y_2 \, p^2 + \cdots)$, where $|y|_p = p^{- l} =
\rho$. For $|x|_p$ there are the following possibilities: (i) $|x|_p
= \rho > |\xi|_p \,,$ if $k > l$  (ii) $|x|_p = |\xi|_p > \rho ,$ if
$k < l$ (iii) $|x|_p = |\xi|_p = \rho$ if $k = l$ and $\xi_0 + y_0
\neq p$, and (iv) $|x|_p < |\xi|_p = \rho$ if $k = l$ and $ \xi_0 +
y_0 = p $. When $\xi$ is fixed  then $|x|_p$ depends on  $\rho .$

 For  more details
about $p$-adic numbers and their algebraic extensions, see, e.g.
\cite{schikhof}.

 To consider real and $p$-adic numbers simultaneously and on the equal
footing one uses concept of adeles. An adele $x$ (see, e.g.
\cite{gelfand}) is an infinite sequence \beq
  x= (x_\infty\,, x_2\,, x_3\,, \cdots, x_p\,, \cdots), \label{2.4}\eeq
where $x_\infty \in {\mathbb R}$ and $x_p \in {\mathbb Q}_p$ with
the restriction that for all but a finite set $\mathcal P$ of primes
$p$ one has  $x_p \in {\mathbb Z}_p $. Componentwise addition and
multiplication make the ring structure of the set ${\mathbb A}$ of
all adeles, which is the union of restricted direct products in the
following form:
\begin{equation}
 {\mathbb A} = \bigcup_{{\mathcal P}} {\mathbb A} ({\mathcal P}),
 \ \ \ \  {\mathbb A} ({\mathcal P}) = {\mathbb R}\times \prod_{p\in
 {\mathcal P}} {\mathbb Q}_p
 \times \prod_{p\not\in {\mathcal P}} {\mathbb Z}_p \, .         \label{2.5}
\end{equation}

A multiplicative group of ideles $\mathbb{A}^\ast$ is a subset of
${\mathbb A}$ with elements $x= (x_\infty\,, x_2\,, x_3 \,, \cdots ,
x_p\,, \cdots)$ ,  where $x_\infty \in {\mathbb R}^\ast = {\mathbb
R} \setminus \{ 0\}$ and $x_p \in {\mathbb Q}^\ast_p = {\mathbb Q}_p
\setminus \{0 \}$ with the restriction that for all but a finite set
$\mathcal P$  one has that  $x_p \in {\mathbb U}_p$ . Thus the whole
set of ideles is
\begin{equation}
 {\mathbb A}^\ast = \bigcup_{{\mathcal P}} {\mathbb A}^\ast ({\mathcal P}),
 \ \ \ \ {\mathbb A}^\ast ({\mathcal P}) = {\mathbb R}^{\ast}\times \prod_{p\in {\mathcal P}}
 {\mathbb Q}^\ast_p
 \times \prod_{p\not\in {\mathcal P}} {\mathbb U}_p \, .         \label{2.6}
\end{equation}

A principal adele (idele) is a sequence $ (x, x, \cdots, x, \cdots)
\in {\mathbb A}$ , where $x \in  {\mathbb Q}\quad (x \in {\mathbb
Q}^\ast = {\mathbb Q}\setminus \{ 0\})$. ${\mathbb Q}$ and ${\mathbb
Q}^\ast$ are naturally embedded in  ${\mathbb A}$ and ${\mathbb
A}^\ast$ , respectively.

\section{\large Linear Fractional Dynamical Systems}

Let us first recall some basic notions from the theory of dynamical
systems \cite{khrennikov4} valid for mapping (\ref{1.1}) and its
iterations (\ref{1.2}) at real and $p$-adic spaces. Let us introduce
an index $v$ to denote real ($v = \infty$) and $p$-adic ($v = p$)
cases simultaneously. A \textit{ fixed point} $\xi$ is a solution of
the equation $f (\xi) = \xi .$ If there exists a neighborhood $V_v
(\xi)$ of the fixed point $\xi$ such that for any point $x_n \in V_v
(\xi), \,\, x_n \neq \xi ,$ holds: $(i) \,\, |x_n -\xi |_v <
|x_{n-1} - \xi|_v$, i.e. $\lim_{n \to \infty} x_n = \xi$, then $\xi$
is called an \textit{attractor}; $(ii) \,\, |x_n -\xi |_v
> |x_{n-1} - \xi|_v$, then $\xi$ is  a \textit{repeller}; and $(iii)
\,\,  |x_n -\xi |_v = |x_{n-1} - \xi|_v $, then $\xi$ is  an
\textit{indifferent point}.
 Basin of attraction $A_v (\xi)$ of an attractor
$\xi$ is the set \beq A_v (\xi) = \{x_0 \in \mathbb{Q}_v : \lim_{n
\to \infty} x_n \to \xi \} .\label{3.1}\eeq A Siegel disk is called
an open ball $V_v (r ,\xi)$ if every sphere $S_v (\rho ,\xi), \,
\rho < r$ is an invariant sphere of the mapping $f (x)$, i.e. if an
initial point $x_0 \in S_v (\rho ,\xi)$ then all iterations $x_n$
also belong to $S_v (\rho ,\xi)$. The union of all Siegel disks $V_v
(r ,\xi)$ with the same center $\xi$ is called a maximum Siegel disk
and denoted by $S I_v (\xi)$. Invariant spheres $S_v (\rho ,\xi_i)$
of Siegel disks $V_v (r ,\xi_i)$ for indifferent fixed points
$\xi_i$ have to satisfy $|x_n - \xi_i|_v = |x_0 - \xi_i|_v = \rho_v
< r_v$ for all $n \in \mathbb{N}$.

  When the mapping (\ref{1.1}) has the first derivative in the fixed point $\xi$
then one can use the following properties: $|f' (\xi)|_v < 1$ -
attractor, $|f' (\xi)|_v > 1$ - repeller and $|f' (\xi)|_v = 1$ -
indifferent point.

We shall mainly consider  rational dynamical systems given by map
(\ref{1.3}) which is isomorphic to the matrix

\bea F =  \left(\begin{array}{ll}
 a  &  b  \\
 c &   d
                \end{array}
                \right)\,,
\quad \mbox{det} F = 1 \,,   \label{3.2}
                \eea

\noi where $a, b, c, d \in \mathbb{Q}$ and with condition  $ a d - b
c = 1$. The corresponding group of matrices $F\,,$ with $\,
\mbox{det}\, F = 1$, is $S L (2, \mathbb{Q})$.

Recall that iteration (\ref{1.2}) may have  periodic points. A point
$x_0$ is called a periodic point if there exists $k$ such that $f^k
(x_0) = x_0$. The smallest such $k$ is the period of $x_0$ and then
$x_0$ is called a $k$-periodic point. Note that fixed points are
$1$-periodic points. Iteration (\ref{1.2}) can be periodic for all
points $x_0 \in X$. Our map (\ref{1.3}) generates periodicity  of a
period $k$ when related matrix (\ref{3.2}) satisfies $F^k = {I}$,
where $I$ is $2 \times 2$ unit matrix. For example, if $d = -a$ and
$a^2  + b \, c = 1$ one has $k = 2$ periodicity.

It is worth mentioning that the map (\ref{1.3}) preserves the
cross-ratio \beq   \frac{(\alpha_1 - \alpha_3) \, (\alpha_2 -
\alpha_4)}{(\alpha_1 - \alpha_4)\, (\alpha_2 - \alpha_3)}  =
  \frac{(f ( \alpha_1) - f (\alpha_3)) \, (f (\alpha_2 ) -
f (\alpha_4) )}{ (f (\alpha_1) - f (\alpha_4) )\, ( f(\alpha_2) - f
(\alpha_3 ))} \label{3.3} \eeq between any different points $x =
\alpha_1\,, \alpha_2\,, \alpha_3\,, \alpha_4\,$.

 To be (\ref{1.3}) an adelic system, it must be satisfied $|f_p (x_p)|_p \leq
1$ in
 \bea
f_{\mathbb{A}}(x) = \Big( f_\infty (x_\infty) \,, f_2 (x_2)\,, f_3
(x_3)\,, \cdots \,, f_p (x_p) \,, \cdots \Big) \,, \quad x \in
\mathbb{A}\,, \label{3.4} \eea

\noi for all but a finite set $\mathcal{P}$ of prime numbers $p$. In
other words, there has to be a prime number $q$ such that $|f_p
(x_p)|_p \leq 1$  for
 all $p > q$. It is shown in \cite{mihajlovic} that function (\ref{1.3})
 satisfies adelic behavior.

   \vskip.4cm


For the function (\ref{1.3}) we find the following two fixed points:

\bea \xi_{1,2} = \frac{a -d \pm \sqrt{(a -d)^2 + 4 a d - 4}}{2c} \,
= \frac{a -d \pm \sqrt{(a + d)^2 - 4}}{2c} \, \label{3.5} \eea with
condition $ad - bc = 1$ and properties \beq f(\xi_1)\cdot f(\xi_2) =
\xi_1 \cdot \xi_2 = - \frac{b}{c} \, \,, \quad \quad f' (\xi_1)
\cdot f' (\xi_2) = 1. \label{3.6} \eeq

For the fixed points it is important to notice that if the point
$\xi_1$ is attractive ($|f' (\xi_1)|_v < 1$) then the point $\xi_2$
is repelling ($|f' (\xi_2)|_v > 1$) and vice versa. The indifferent
fixed points always emerge in the pair. These facts obviously follow
from the relation (\ref{3.6}).
 Generally, these
points belong to $\mathbb{C}$ in real case and $\mathbb{C}_p$ in
$p$-adic case, and their analysis will be done elsewhere.

We are interested in rational fixed points because they
simultaneously belong to real and $p$-adic numbers.  Fixed rational
points (\ref{3.5}) for the dynamical system (\ref{1.3}) have been
investigated in \cite{mihajlovic} for the following four particular
cases: ({\bf A}) $b= 0$, ({\bf B}) $ b = c, \, d = a $, ({\bf C}) $
d = - a + 2$ and ({\bf D}) $ d = - a - 2$. Basins of attraction, the
Siegel disks and adelic trajectories are examined for the case ({\bf
A}).

In this paper we continue investigation started in
\cite{mihajlovic}. First of all let us note that the general case of
rational fixed points, i.e.

\beq (a + d)^2 - 4 = \delta^2 \,, \quad a\, d - b\, c = 1\,, \quad
\delta \in \mathbb{Q}  \label{3.9} \eeq has solution.  Namely, the
hyperbolic  equation $(a + d)^2 - 4 = \delta^2$ has rational
solution in the form

\beq
 a + d = \pm \, \frac{2\, (1 + t^2)}{1 - t^2}\,, \quad \delta = \frac{4\, t}{1 -
 t^2}\,,  \quad t \in \mathbb{Q}\setminus \{1\,, -1 \}\,. \label{3.10}
\eeq  For given parameters $a,\, d$ and $\delta$ one has $t = (a + d
\pm 2)/ \delta$. Rational values for parameters $a$ and $b$ follow
from the expression (\ref{3.10}). Then $b$ and $c$ are also
rational, because

\beq
 b\, c = \frac{\delta^2 - a^2 - d^2}{2} + 1\,,  \quad c \neq 0\,.
 \label{3.11}
\eeq Three of these parameters $a,\, b,\, c,\, d\, \, \mbox{and}
\,\, t$ are free. In the cases ({\bf A}) and ({\bf B}) parameter $t
\in \mathbb{Q}\setminus \{1\,, -1 \}$. For an analysis of the cases
({\bf A}) and ({\bf B}) see Ref. \cite{mihajlovic}. Now we are going
to investigate the cases  ({\bf C}) and ({\bf D}) in more details.
These two cases exhaust all possibilities with $\delta = t = 0$.

\subsection{Case (C): $\delta = t = 0 , \, d = - a + 2 , \, (a - 1 )^2 + b c = 0$.}

This is a case with double fixed point:

\beq
 f (x) = \frac{a x + b}{c x - a + 2}\,, \quad  \xi_1 = \xi_2 =\frac{a-1}{ c}
 \,. \label{3.22}
\eeq For further investigation we need
 \bea f' (x) = \frac{1}{(c x
-a + 2)^2} \,, \quad f'(\xi_1) =f'(\xi_2)= 1 \,. \label{3.23} \eea

 Due to $|f' (\xi_1)|_v = |f' (\xi_2)|_v =1$ it follows that the
fused rational fixed point  $\xi_1 = \xi_2 = \frac{a-1}{ c} $ is
indifferent one in real as well as in all $p$-adic cases.

According to the above results we have only one adelic fixed point
$\xi^{(1)} = \xi^{(2)} \equiv \xi$, i.e.

\bea \xi = \big(\xi_\infty \,, \xi_2\,,\xi_3\,,\xi_5\,, \cdots \,,
\xi _p \,, \cdots \Big) \,, \quad \xi \in \mathbb{A}\,, \label{3.24}
\eea where $\xi_\infty = \xi_p = \frac{a-1}{ c}$  for any $p$. This
is one pure adelic indifferent point for any rational values of
parameters $a\,, b$ and $c$ constrained by relation $ (a - 1)^2 +
b\, c =0 $ and $c \neq 0$.

The $n$-th iteration is

\beq x_n = \frac{(n\, a - n + 1)\, x_0  + n\, b}{ n\, c\, x_0  - n\,
a + n + 1} \,, \eeq where $x_0$ is an initial state. In the real
case for all $x_0 \neq (n\, a -n-1)/ n c$ we have $x_n \to \frac{a -
1}{c}$ when $n \to \infty .$

\subsubsection{ Subcase: \, $b = - c \,, \,\, d = a - 2 c \,, \,\,  (a - c)^2 = 1$.}

In this case one has again mapping with fused fixed points, i.e.

\beq
 f (x) = \frac{a x - c}{c x + a - 2c}\,, \quad  \xi_1 = \xi_2 = 1
 \,. \label{3.19}
\eeq In the following we need

\bea f' (x) = \frac{1}{(c x + a - 2c)^2} \,, \quad f'(\xi_1)
=f'(\xi_2)= 1\,. \label{3.20} \eea

In this special case we have the only one possibility.  Namely, due
to $|f' (\xi_1)|_v = |f' (\xi_2)|_v =1$ it follows that the fused
fixed point  $\xi_1 = \xi_2 = 1 $ is indifferent one in real as well
as in all $p$-adic cases.

According to the above results one has only one adelic fixed point
$\xi^{(1)} = \xi ^{(2)} \equiv \xi$, i.e.

\bea \xi = \big(\xi_\infty \,, \xi_2\,,\xi_3\,,\xi_5\,, \cdots \,,
\xi_p \,, \cdots \Big) \,, \quad \xi \in \mathbb{A}\,, \label{3.21}
\eea where $\xi_\infty = \xi_p =  1$  for any $p$. This is one pure
adelic indifferent point for any rational values of parameters $a$
and $c$ constrained by relation $(a -c)^2 = 1$ and $c \neq 0$.

The $n$-th iteration is

\beq x_n = \frac{[a + (n - 1) c]\, x_0  - n c}{ n c \, x_0 + a - (n
+ 1) c } \,, \eeq which in the real case gives $x_n \to 1$ when $n
\to \infty$ and $x_0 \neq \frac{(n + 1)\, c  - a}{n \, c}$.

\subsection{Case (D): $\delta = t = 0 , \, d = - a - 2 , \, (a + 1)^2 + b c = 0$.}

As in the previous  case one has here coincidence  of fixed points.
Namely,

\beq
 f (x) = \frac{a x + b}{c x - a - 2}\, \quad  \xi_1 = \xi_2 =\frac{a + 1}{ c} \,.
\label{3.25}\eeq  We also employ
 \bea f' (x) = \frac{1}{(c x
-a - 2)^2} \,, \quad f'(\xi_1) =f'(\xi_2)= 1 \,. \label{3.26} \eea

Since $|f' (\xi_1)|_v = |f' (\xi_2)|_v =1$ it follows that the fused
fixed point  $\xi_1 = \xi_2 = \frac{a+1}{ c} $ is indifferent one in
real as well as in all $p$-adic cases.

From the above results one has only one adelic fixed point
$\xi^{(1)} = \xi^{(2)} \equiv \xi$, i.e.

\bea \xi = \big(\xi_\infty \,, \xi_2\,,\xi_3\,,\xi_5\,, \cdots \,,
\xi _p \,, \cdots \Big) \,, \quad \xi \in \mathbb{A}\,, \label{3.27}
\eea where $\xi_\infty = \xi_p = \frac{a+1}{ c}$  for any $p $. This
is one pure adelic indifferent point for any rational values of
parameters $a$ and $c$ constrained by relation $ (a + 1 )^2 + b\,c =
0 $ and $c \neq 0$.

The $n$-th iteration is

\bea x_n &=& \frac{ (-1)^{n +1} [n \, (a + 1)  - 1 ]\, x_0 + (-1)^{n
+1}\, n\, b}{ (-1)^{n +1}\, n\, c\, x_0 - (-1)^{n +1}\, [ n\, (a +
1) + 1 ]  } \,, \eea where $x_0 \neq \frac{n\, (a +1) + 1}{n\, c}$
is an initial state. In the real case $x_n \to \frac{a +1}{c}$ when
$n \to \infty$.

\subsubsection{ Subcase: \, $b = - c \,, \,\, d = a + 2 c \,, \,\,  (a + c)^2 = 1$.}

This is the case with fused fixed points

\beq
 f (x) = \frac{a x - c}{c x + a + 2c}\,, \quad  \xi_1 = \xi_2 = -1 \,.
\label{3.16}\eeq For further investigation we need
 \bea f' (x) = \frac{1}{(c x +
a + 2c)^2} \,, \quad f'(\xi_1) =f'(\xi_2)= 1 \,. \label{3.17} \eea

In this special case we have the only one possibility.  Namely, due
to $|f' (\xi_1)|_v = |f' (\xi_2)|_v =1$ it follows that the fused
fixed point  $\xi_1 = \xi_2 = -1 $ is indifferent one in real as
well as in all $p$-adic cases (i.e. for all primes $p$).

According to the above results one has only one adelic fixed point
$\xi^{(1)} = \xi^{(2)} \equiv \xi$, i.e.

\bea \xi = \big(\xi_\infty \,, \xi_2\,,\xi_3\,,\xi_5\,, \cdots \,,
\xi _p \,, \cdots \Big) \,, \quad \xi \in \mathbb{A}\,, \label{3.18}
\eea where $\xi_\infty = \xi_p = - 1$  for any $p$. This is one pure
adelic indifferent point for any rational values of parameters $a$
and $c$ constrained by relation $(a +c)^2 = 1$ and $c \neq 0$.

The $n$-th iteration is

\beq x_n = \frac{[a - (n - 1) c]\, x_0  - n c}{ n c \, x_0 + a + (n
+ 1) c } \,, \eeq which for $x_0 \neq - \,\frac{(n + 1)\, c + a}{n\,
c}$ leads to $x_n \to - 1$ when $n \to \infty$ in the real case.

\section{Concluding Remarks}

According to \cite{mukhamedov1} radius of the $p$-adic Siegel disks
in all above considered cases is $r = \frac{|a|_p}{|c|_p}$.

In the above analysis the space of states X can be extended to the
whole projective line {\bf P}$^1$. Then $x_0 = -d/c$  maps to the
point at infinity and $x_0 = \infty$ maps to $a/c$.

There are many possibilities for generalization of dynamical system
(\ref{1.3}). Two directions seem to be very interesting: 1) maintain
one-dimensional space of states and increase nonlinearity and 2)
maintain nonlinearity but increase dimensionality of the space of
states.

Under 1) we understand $f (x) = P_k (x)/ Q_l (x)$, where $P_k (x)$
and $Q_l (x)$ are polynomials of degrees $k$ and $l$, respectively.
In particular, one can take $f (x) = \prod_{i = 1}^k (a_i \, x +
b_i)/ (c_i \, x + d_i)$ with some restrictions on parameters $a_i\,,
b_i\,, c_i$ and $d_i$ . Already $f (x) = \prod_{i = 1}^2 (a_i \, x +
b_i)/ (c_i \, x + d_i)$ contains some intresting cases (see
\cite{mukhamedov2} and references therein).

Direction 2) has the form $f_i (x) = ( \sum_{ j =1}^k \alpha_{i j}\,
x_j + \alpha_{i 0})/ ( \sum_{ j =1}^k \beta_{i j}\, x_j + \beta_{i
0})$, where $i = 1 ,\, 2 ,\, \cdots , k$. Two-dimensional case also
offers a rich structure. For instance, iterative projective
transformations

\bea x_n = \frac{a_{1 1} \, x_{n - 1} + a_{1 2}\, y_{n -1} + a_{1 3}
}{a_{3 1} \, x_{n-1}
+ a_{3 2}\, y_{n -1} + a_{3 3} } \,,\\
 y_n = \frac{a_{2 1} \, x_{n -1} + a_{2 2}\, y_{n -1} + a_{2 3} }{a_{3 1} \,
 x_{n -1} +  a_{3 2}\, y_{n -1} + a_{3 3} }  \eea

are isomorphic to matrices

\bea F =  \left(\begin{array}{lll}
 a_{1 1}  &  a_{1 2} & a_{1 3}  \\
 a_{2 1} &   a_{2 2} & a_{2 3}  \\
 a_{3 1}  &  a_{3 2} & a_{3 3}
                \end{array}
                \right)\,, \quad \mbox{det } F \neq 0 \,.
\eea

Another possibility is to consider the recurrence relation
\cite{bedford}

\beq x_{n + k +1} = \frac{\alpha_0 + \alpha_1 \, x_{n +1} + \cdots +
\alpha_k \, x_{n + k}}{\beta_0 + \beta_1 \, x_{n +1} + \cdots +
\beta_k \, x_{n +k}} \,, \eeq where $\alpha_0, \cdots , \alpha_k$
and $\beta_0, \cdots , \beta_k$ are given rational numbers. Here an
initial $k$-tuple $(x_1\,, \cdots \,, x_k)$ generates an infinite
sequence of states by map

\beq f (x_1\,, \cdots \,, x_k) = \Big( x_2\,, \cdots \,, x_k \,, \,
\frac{\alpha_0 + \alpha_1 \, x_{1} + \cdots + \alpha_k \,
x_{k}}{\beta_0 + \beta_1 \, x_{1} + \cdots + \beta_k \, x_{k}} \Big)
\,, \eeq for which  periodicity of the case $k = 2$ is investigated
in \cite{bedford}.
\bigskip

\bigskip

\section*{\large Acknowledgments}

The work on this article was partially supported by the Ministry of
Science and Environmental Protection, Serbia, under contract No
144032D.

\end{document}